\documentclass[12pt]{article}
\usepackage{amsfonts}
\usepackage{color}
\newcommand{\Section}[1]%
{\section{#1}\setcounter{equation}{0}%
\setcounter{theorem}{0}}
\newtheorem{theorem}{Theorem}

\def\re{\mathbb{R}}
\def\co{\mathbb{C}}
\def\av{\mathbb{E}}
\def\ze{\mathbb{Z}}

\newenvironment{proof}[1]%
{\par\noindent{\em #1:\ }}%
{~\rule{2mm}{2mm}\par\bigskip}
\begin{document}
%%%%%%%%%%%%%%%%%%%%%%%%%%%%%%%%%%%%%%%%%%%
%%%%%%%%%%%%%%%%%%%%%%%%%%%%%%%%%%%%%%%%%%%
%%%%%%%%%%%%%%%%%%%%%%%%%%%%%%%%%%%%%%%%%%%
\newpage\thispagestyle{empty}
{\topskip 2cm
\begin{center}
{\Large\bf Two No-Go Theorems on Superconductivity\\} 
\bigskip\bigskip
{\Large Yasuhiro Tada\footnote{\small \it Institute for Solid State Physics, The University of Tokyo, 
Kashiwa 277-8581, JAPAN, 
{\small\tt e-mail: tada@issp.u-tokyo.ac.jp}}$^{,}$
\footnote{\small \it 
Max Planck Institute for the Physics of Complex Systems,
N{\"o}thnitzer Str. 38, 01187 Dresden, GERMANY}
and Tohru Koma\footnote{\small \it Department of Physics, Gakushuin University, Mejiro, Toshima-ku, Tokyo 171-8588, JAPAN,
{\small\tt e-mail: tohru.koma@gakushuin.ac.jp}}
\\}
%\bigskip
%\bigskip
\end{center}
%This is also for double spacing
%\newpage
\vfil
\noindent
We study lattice superconductors such as attractive Hubbard models. 
As is well known, Bloch's theorem asserts absence of persistent current 
in ground states and equilibrium states for general fermion systems. 
While the statement of the theorem is true, 
we can show that the theorem cannot exclude 
possibility of a surface persistent current. 
Such a current can be 
stabilized by boundary magnetic fields
which do not penetrate into the bulk region of a superconductor, 
provided emergence of massive photons, i.e., Meissner effect. 
Therefore, we can expect that a surface persistent 
current is realized for a ground/equilibrium state in the sense of stability against local perturbations. 
We also apply Elitzur's theorem to superconductors at finite temperatures. 
As a result, we prove absence of symmetry breaking of the global $U(1)$ phase of electrons 
for almost all gauge fields. These observations suggest that the nature of superconductivity 
is the emergence of massive photons rather than the symmetry breaking of the $U(1)$ phase of electrons.      
\par
%%%%%%%%%%%%%%%%%%%%%%%%%%%%%%%%%%%%%%%%%%%%%%%%%%%%%%%%%%%%%%%%%%%%%%%%%%%
\noindent
\bigskip
\hrule
\bigskip
\noindent
{\bf KEY WORDS:} Superconductivity, Persistent current, $U(1)$ symmetry breaking, Bloch's theorem, 
Elitzur's theorem, Meissner effect\hfill
%%%%%%%%%%%%%%%%%%%%%%%%%%%%%%%%%%%%%%%%%%%%%%%%%%%%%%
\vfil}
%\newpage
%%%%%%%%%%%%%%%%%%%%%%%%%%%%%%%%%%%%%%%%%%%

%%%%%%%%%%%%%%%%%%%%%%%%%%%%%%%%%%%%%%%%%%%%%%%%%%%%%%
%%%%%%%%%%%%%%%%%%%%%%%%%%%%%%%%%%%%%%%%%%%
\Section{Introduction}
\label{Introduction}

Superconductivity still has some mysteries to solve although it has been one of the central 
issues in condensed matter physics.
Bloch's theorem \cite{Bohm,Sewell,Vignale,OM,KO,Tada,Yamamoto} states absence of 
bulk persistent current in ground states and equilibrium states for general fermion systems, 
whereas experiments of superconductors \cite{KHS} show 
persistent current with toroidal geometry.   
Elitzur's theorem \cite{Elitzur,DdFG,BN2} states that local gauge symmetries including 
quantum mechanical $U(1)$ symmetry cannot be broken spontaneously, 
whereas superconductivity has been often explained by using the idea that  
the $U(1)$ symmetry is spontaneously broken with a nonvanishing expectation value of an order 
parameter such as the Cooper pair {amplitude}. 
Although these two theorems have a long history, 
it is still unclear how the two theorems are reconciled with the conventional theory of superconductivity. 

If the superconducting state which carries the persistent current in the experiments 
is  neither a ground state nor a thermal equilibrium state, 
then the state must be an excited state or a nonequilibrium state which has a very long lifetime. 
Such a state is often said to be metastable \cite{Sewell}. What mechanism makes the lifetime of 
the metastable state carrying the persistent current very long? 
One of the important stuffs is multiply-connected geometry
such as toroid. In fact, we can expect that, when magnetic fields are wound up the surface
of the large toroid in a stable state which is realized by Meissner
effect, changing the topology of the global structure
needs a large energy which is determined by the system size.

In the present paper, we consider possibility of a ground/equilibrium state which carries a persistent current, 
and show that Bloch's theorem cannot exclude such a state if 
the persistent current is localized near the surface of the sample 
with boundary magnetic fields which are stabilized by Meissner effect. 
Therefore, we can expect that a surface persistent 
current is realized for {\it a ground/equilibrium state} in the sense of stability against local perturbations.
Such a current carrying state has a long lifetime even in a finite size system, 
because a transition from the corresponding state to other states costs larger and larger energy  
as the system size increases. 

In order to explain Meissner effect in the above argument, 
one has to deal with the interactions between electrons and electro-magnetic fields.
As mentioned above, however, Elitzur's theorem \cite{Elitzur} 
implies that the {local} $U(1)$ gauge symmetry cannot be broken spontaneously.  
Actually, we can prove, by relying on the argument of \cite{Elitzur}, that the fluctuations of the gauge fields yield 
absence of the $U(1)$ symmetry breaking for superconductors at finite temperatures.  
Surprisingly, even for the systems with a fixed configuration of gauge fields, 
we can prove absence of symmetry breaking of the global $U(1)$ phase of electrons 
for almost all configurations of gauge fields in the sense of macroscopic spontaneous magnetization. 
This is slightly different from the former consequence of Elitzur's theorem. 

These observations in the present paper strongly suggest that the nature of superconductivity 
is the emergence of massive photons rather than the symmetry breaking of the $U(1)$ phase of electrons.  
In fact, it is known that, in Higgs models coupled with gauge fields \cite{BN2,FMS,BN1,{Seiler}}, 
a transition from massless photons to massive photons is possible for varying the coupling constant, 
irrespective of whether or not the $U(1)$ symmetry breaking occurs.

The present paper is organized as follows: In Sec.~\ref{Bloch}, we give the precise definition of 
the models which we consider, and examine Bloch's theorem in a mathematically rigorous manner. 
As a result, we show that Bloch's theorem cannot exclude the possibility of surface persistent current.   
In Sec.~\ref{sec:Elitzur}, we apply Elitzur's theorem to two situations, the present models with  
annealed and quenched gauge fields. In both cases, we prove absence of the $U(1)$ symmetry breaking. 

%%%%%%%%%%%%%%%%%%%%%%%%%%%%%%%%%%%%%%%%%%%%%%%%
\Section{Absence of bulk persistent current} 
\label{Bloch}

In this section, we show that Bloch's theorem does not exclude possibility of 
a surface current in a ground/equilibrium state, by examining the proof in a mathematically 
rigorous manner. 

\subsection{Models}

Consider a $d$-dimensional connected graph $G:=(\Lambda,{\cal B})$, 
where $\Lambda$ is a set of lattice sites and ${\cal B}$ is a set of bonds, i.e., pairs of lattice sites.  
The graph-theoretic distance, ${\rm dist}(x,y)$, between two lattice sites $x,y$ is defined as the minimum 
number of bonds in ${\cal B}$ that one needs to connect $x$ and $y$.  

We consider a tight-binding model on the lattice $\Lambda$.
The Hamiltonian $H_\Lambda$ is given by 
\begin{equation}
H_\Lambda:=\sum_{x,y\in\Lambda}\sum_{\alpha,\beta}t_{x,y}^{\alpha,\beta}c_{x,\alpha}^\dagger c_{y,\beta}
+\sum_{I\ge 1}\sum_{x_1,\sigma_1}\sum_{x_2,\sigma_2}\cdots\sum_{x_I,\sigma_I}
W_{x_1,\sigma_1;x_2,\sigma_2;\ldots;x_I,\sigma_I}n_{x_1,\sigma_1}n_{x_2,\sigma_2}\cdots n_{x_I,\sigma_I},
\label{HamLambda}
\end{equation}
where $c_{x,\sigma}^\dagger,c_{x,\sigma}$ are, respectively, the creation and annihilation fermion operators 
at the site $x\in\Lambda$ 
with the internal degree of freedom, $\sigma$, such as spin or orbital; 
the hopping amplitudes $t_{x,y}^{\alpha,\beta}$ are complex numbers which satisfy 
the Hermitian conditions, 
$$
t_{y,x}^{\beta,\alpha}=\left(t_{x,y}^{\alpha,\beta}\right)^\ast,
$$
and the coupling constants $W_{x_1,\sigma_1;x_2,\sigma_2;\ldots;x_I,\sigma_I}$ of the interactions are 
real numbers. As usual, we have written $n_{x,\sigma}=c_{x,\sigma}^\dagger c_{x,\sigma}$ 
for the number operators of the fermion with $\sigma$ at the site $x$. We assume that 
both of the hopping amplitudes and the interactions are of finite range in the sense of the graph-theoretic 
distance, and assume that all of the strengths are uniformly bounded as 
$$
\left|t_{y,x}^{\beta,\alpha}\right|\le t_0 \quad\mbox{and}\quad 
\left|W_{x_1,\sigma_1;x_2,\sigma_2;\ldots;x_I,\sigma_I}\right|\le W_0 
$$
with some positive constants, $t_0$ and $W_0$. 

%%%%%%%%%%%%%%%%%%%%%%%%%%%%%%%%%%%%%%%%%%%%%%%%%%%%%%%
\subsection{Current operators}

Consider a cylindrical region $\Omega$ which is written a disjoint union of 
$(d-1)$-dimensional discs $D_j$ with a common radius $R$ as  
$$
\Omega=D_0\cup D_1\cup D_2\cup\cdots \cup D_L\subset\Lambda 
$$
with a large positive integer $L$. 
Namely, the region $\Omega$ consists of $L$-layers of the segments $D_j$ of the $(d-1)$-dimensional 
hyperplanes. We assume that the distance between two discs $D_i$ and $D_j$ satisfies 
$$
{\rm dist}(D_i,D_j)=|i-j|
$$
for $i,j=1,2,\ldots,L$. 

In order to define the current operator on the region $\Omega$, we consider 
a larger region $\tilde{\Omega}$ which is written a disjoint union of large discs $\tilde{D}_j$ as 
$$
\tilde{\Omega}=\tilde{D}_{-\tilde{L}}\cup\tilde{D}_{-\tilde{L}+1}\cup\cdots\cup
\tilde{D}_{-1}\cup\tilde{D}_0\cup\tilde{D}_1\cup\tilde{D}_2\cup\cdots\cup\tilde{D}_{\tilde{L}-1}
\cup\tilde{D}_{\tilde{L}}\subset\Lambda,
$$
where $\tilde{L}$ is a large positive integer satisfying $\tilde{L}>L$. 
We choose the discs $\tilde{D}_j$ so that the following conditions satisfy: 
$$
\tilde{D}_j\supset D_j \quad \mbox{for \ } j=0,1,2,\ldots,L,
$$ 
$$
{\rm dist}(\tilde{D}_i,\tilde{D}_j)=|i-j|\quad 
\mbox{for \ } i,j=-\tilde{L},-\tilde{L}+1,\ldots,-1,0,1,2,\ldots,\tilde{L}-1,\tilde{L},
$$
and 
$$
{\rm dist}(\Omega,\partial\tilde{\Omega})\ge r_0, 
$$
where the boundary $\partial\tilde{\Omega}$ of the region $\tilde{\Omega}$ is given by 
$$
\partial\tilde{\Omega}:=\{x\in\tilde{\Omega}\;|\;{\rm dist}(x,\Lambda\backslash \tilde{\Omega})=1\}
$$
and $r_0$ is the maximum hopping range, i.e.,  
$$
r_0:=\max\{{\rm dist}(x,y)\; |\; t_{x,y}^{\alpha,\beta}\ne 0\}.
$$
Clearly, one has $\Omega\subset\tilde{\Omega}$. 

Let $\varphi\in\ell^2(\Lambda,{\co}^M)$ be a wavepacket of single fermion, where ${\co}^M$ is 
the space of the internal degree of freedom with a finite dimension $M$. 
We assume that the support of the wavepacket $\varphi$ 
is contained in the region $\tilde{\Omega}$. 
We define the position operator $X$ as 
$$
(X\varphi)(x)=j\varphi(x)\quad \mbox{for \ } x\in \tilde{D}_j. 
$$
We introduce the step functions,  
$$
\vartheta_j(x):=\cases{1, & $x\in\tilde{\Omega}_{[j,\tilde{L}]}$;\cr
0, & otherwise,}
$$
with the kink $\tilde{D}_j$, where we have written 
$$
\tilde{\Omega}_{[j,\tilde{L}]}:=\tilde{D}_j\cup\tilde{D}_{j+1}\cup\cdots\cup\tilde{D}_{\tilde{L}}. 
$$
Then, the position operator $X$ is written as 
$$
X=\sum_{j=-\tilde{L}+1}^{\tilde{L}}\vartheta_j-\tilde{L}
$$
in terms of the step functions. We write $H_{\tilde{\Omega}}^{(1)}$ for the single-fermion Hamiltonian 
on the region $\tilde{\Omega}$. The current operator is given by \cite{Mahan,Koma} 
$$
J^{(1)}:=i[H_{\tilde{\Omega}}^{(1)},X].
$$
{From} the above expression of the position operator $X$, we have 
$$
J^{(1)}:=i[H_{\tilde{\Omega}}^{(1)},X]
=\sum_{j=-\tilde{L}+1}^{\tilde{L}}i[H_{\tilde{\Omega}}^{(1)},\vartheta_j].
$$
for the above wavepacket $\varphi$. Since the summand in the right-hand side is nothing but 
the current density operator across the disc ${\tilde D}_j$, we write 
$$
J_j^{(1)}:=i[H_{\tilde{\Omega}}^{(1)},\vartheta_j]
$$
for the operator. In order to obtain the corresponding current density operator for many fermions, 
we introduce the step function as  
$$
\Theta_j:=\sum_\sigma \sum_{x\in\tilde{\Omega}_{[j,{\tilde L}]}} n_{x,\sigma}
$$
by using the number operators $n_{x,\sigma}$ of fermions. 
In consequence, the current density operator across the disc ${\tilde D}_j$ for many fermions 
is written 
\begin{equation}
J_j:=i[H_{\tilde{\Omega}},\Theta_j]
\end{equation}
for the wavefunction $\Phi$ which is supported by the region $\tilde{\Omega}$, where 
the Hamiltonian $H_{\tilde{\Omega}}$ is the restriction of the Hamiltonian $H_\Lambda$ of (\ref{HamLambda}) 
to the region $\tilde{\Omega}$. 
By using the expression (\ref{HamLambda}) of the Hamiltonian $H_\Lambda$, we obtain the explicit form of 
the operator $J_j$ as 
$$
J_j=\sum_{\alpha,\beta}\sum_{x\in\tilde{\Omega}_{[-\tilde{L},j-1]}}
\sum_{y\in\tilde{\Omega}_{[j,\tilde{L}]}}
i(t_{x,y}^{\alpha,\beta}c_{x,\alpha}^\dagger c_{y,\beta}-t_{y,x}^{\beta,\alpha}c_{y,\beta}^\dagger c_{x,\alpha}),
$$ 
where 
$$
\tilde{\Omega}_{[-\tilde{L},j-1]}:=\tilde{\Omega}\backslash \tilde{\Omega}_{[j,\tilde{L}]}.
$$
Clearly, the above summand is a local current density. Relying on this expression, 
we define the current operator $J_\Omega$ per volume on the region $\Omega$ as  
\begin{equation}
\label{JOmega}
J_\Omega:=\frac{1}{|\Omega|}\sum_{j=1}^L\sum_{\alpha,\beta}\sum_{x\in\Omega_{[0,j-1]}}
\sum_{y\in\tilde{\Omega}_{[j,\tilde{L}]}}
i(t_{x,y}^{\alpha,\beta}c_{x,\alpha}^\dagger c_{y,\beta}-t_{y,x}^{\beta,\alpha}c_{y,\beta}^\dagger c_{x,\alpha}),
\end{equation}   
where $|\cdots|$ stands for the number of the elements in the set, and we have written 
$$
\Omega_{[0,j]}:=D_0\cup D_1\cup \cdots \cup D_j \quad \mbox{for \ } j=0,1,\ldots,L-1.
$$

%%%%%%%%%%%%%%%%%%%%%%%%%%%%%%%%%%%%%%%%%%%%%%%%%%%%%%%%%%%%%%%%%%
\subsection{Bloch's theorem}
  
In order to realize persistent current in the ground state of the present model, we consider  
a toroidal geometry for the graph as an example. 
We impose a boundary condition which mimics the plus boundary condition for ferromagnetic Ising models 
leading to the symmetry breaking phases with plus spontaneous magnetization.
To be specific, we apply a magnetic field tangential to the surface $\partial\Lambda$ of 
the toroidal lattice $\Lambda$ so that surface current perpendicular to the magnetic field 
can appear along the boundary of the toroid. 

Clearly, the total number operator $\sum_{\sigma}\sum_{x\in\Lambda}n_{x,\sigma}$ of the fermions commutes with 
the Hamiltonian $H_\Lambda$. We consider the $N$-fermion ground state $\Phi_{\Lambda,0}^{(N)}$ of 
the Hamiltonian $H_\Lambda$ with the above boundary condition.  
The ground-state expectation is given by 
$$
\omega_{\Lambda,0}^{(N)}(\cdots):=\left\langle\Phi_{\Lambda,0}^{(N)},(\cdots)\Phi_{\Lambda,0}^{(N)}\right\rangle 
$$
with the norm $\Vert\Phi_{\Lambda,0}^{(N)}\Vert=1$.  
We take the infinite-volume limit $|\Lambda_i|\rightarrow\infty$ by using the sequence of 
the lattice $\Lambda_i$ and of the total number $N_i$ of the fermions so that 
the density $\Lambda_i/N_i$ of the fermions converges to a positive constant $\rho$ as $i\rightarrow +\infty$, 
and that the expectation value $\omega_{\Lambda,0}^{(N)}(a)$ converges to the expectation value $\omega_0(a)$ 
with respect to the infinite-volume ground state $\omega_0$ for all the local observables $a$ which we consider. 

The statement of Bloch's theorem is as follows: 

\begin{theorem} The expectation value of the current operator $J_\Omega$ per volume is vanishing 
for the infinite-volume ground state $\omega_0$, i.e., 
\begin{equation}
\label{zeroAvJ}
\lim_{L\nearrow\infty}\lim_{R\nearrow\infty}\omega_0(J_\Omega)=0, 
\end{equation}
where $\Omega$ is the cylindrical region with length $L$ and radius $R$. 
\end{theorem}

\noindent
{\it Remark:} (i) For the two or higher dimensional systems, 
the order of the double limit in (\ref{zeroAvJ}) is not interchangeable. 
In fact, for a fixed $R$, the upper bound of the expectation value in the limit $L\nearrow\infty$ 
becomes infinity and meaningless in the proof below. 
If we want to measure the strength of surface current which is localized near the surface of the sample, 
we must take $R$ to be finite so that the support $\Omega$ of the current operator $J_\Omega$ is localized 
near the surface. Then, Bloch's theorem tells us absolutely nothing about the surface current. 
\smallskip
 
\noindent (ii) Clearly, in one dimensional systems, the surface current itself is meaningless, and 
the net current along the direction of the chain always vanishes, in contrast to higher dimensional systems. 
This is a consequence of lack of a mechanism which stabilizes the current against local perturbations. 
In other words, any current is destroyed by local perturbations in one dimension.
\smallskip

\noindent 
(iii) The boundary magnetic fields are expected to be  
realized by Meissner effect, and once it is generated,
the surface current remains stable in a ground states, i.e., a persistent current.
\smallskip

\noindent
(iv) The extension of the statement of Bloch's theorem to the systems at finite temperatures is straightforward 
by relying on ``passivity" which is a stability property of thermal equilibrium states. 
More precisely, an infinite-volume state $\omega$ is said to be passive if 
$$
\omega(U^\ast[H,U])\ge 0
$$
for any local unitary operator $U$, where $H$ is the Hamiltonian. As is well known, 
all of thermal equilibrium states are passive \cite{BR}.
This fact was pointed out to us by Hal Tasaki.  
\medskip

\begin{proof}{Proof}
We introduce two local unitary transformations,  
\begin{equation}
U_\Omega^{(\pm)}:=\prod_\sigma \prod_{x\in \Omega}\exp\left[i\theta_\pm(x) n_{x,\sigma}\right],
\end{equation}
where 
\begin{equation}
\label{theta}
\theta_\pm(x):=\pm \frac{{2\pi }\ell(x)}{L}
\end{equation}
with 
\begin{equation}
\label{def:ell}
\ell(x):=\cases{j, & $x\in D_j$ for $j=1,2,\ldots,L$; \cr 
                0, & otherwise. \cr}
\end{equation} 
Using these local transformations, we set  
$$
\omega_{\Lambda,\pm}^{(N)}(\cdots)
:=\left\langle\Phi_{\Lambda,0}^{(N)},(U_\Omega^{(\pm)})^\ast(\cdots)
U_\Omega^{(\pm)}\Phi_{\Lambda,0}^{(N)}\right\rangle. 
$$
Since these two states are a local perturbation for the ground state, we have
\begin{equation}
\label{DeltaE} 
\omega_{\Lambda,\pm}^{(N)}(H_\Lambda)-\omega_{\Lambda,0}^{(N)}(H_\Lambda)
=\omega_{\Lambda,0}^{(N)}([(U_\Omega^{(\pm)})^\ast H_\Lambda U_\Omega^{(\pm)}-H_\Lambda])\ge 0. 
\end{equation}
Note that 
$$
\exp[-i\theta_\pm(x)n_{x,\alpha}]c_{x,\alpha}^\dagger \exp[i\theta_\pm(x)n_{x,\alpha}]
=\exp[-i\theta_\pm(x)]c_{x,\alpha}^\dagger
$$
and 
$$
\exp[-i\theta_\pm(y)n_{y,\beta}]c_{y,\beta}\exp[i\theta_\pm(y)n_{y,\beta}]
=\exp[i\theta_\pm(y)]c_{y,\beta}
$$
By using these relations and the expression (\ref{HamLambda}) of the Hamiltonian, 
one has 
$$
(U_\Omega^{(\pm)})^\ast H_\Lambda U_\Omega^{(\pm)}-H_\Lambda
=\sum_{\alpha,\beta}\sum_{x,y}t_{x,y}^{\alpha,\beta}
\left\{\exp[-i(\theta_\pm(x)-\theta_\pm(y))]-1\right\}c_{x,\alpha}^\dagger c_{y,\beta}. 
$$
The right-hand side can be decomposed into three parts as 
\begin{equation}
\label{diffH}
(U_\Omega^{(\pm)})^\ast H_\Lambda U_\Omega^{(\pm)}-H_\Lambda
=\Delta H_\Omega +\Delta H_{\partial\Omega}^{(1)} +\Delta H_{\partial\Omega}^{(2)},
\end{equation}
where 
$$
\Delta H_\Omega=\sum_{\alpha,\beta}\sum_{x,y\in\Omega}t_{x,y}^{\alpha,\beta}
\left\{\exp[-i(\theta_\pm(x)-\theta_\pm(y))]-1\right\}c_{x,\alpha}^\dagger c_{y,\beta},
$$
$$
\Delta H_{\partial\Omega}^{(1)}=\sum_{\alpha,\beta}\sum_{x\in\Omega}\sum_{y\notin\Omega}t_{x,y}^{\alpha,\beta}
\left\{\exp[-i\theta_\pm(x)]-1\right\}c_{x,\alpha}^\dagger c_{y,\beta},
$$
and 
$$
\Delta H_{\partial\Omega}^{(2)}=\sum_{\alpha,\beta}\sum_{x\notin\Omega}\sum_{y\in\Omega}t_{x,y}^{\alpha,\beta}
\left\{\exp[i\theta_\pm(y)]-1\right\}c_{x,\alpha}^\dagger c_{y,\beta},
$$
where we have used (\ref{theta}) and (\ref{def:ell}). 

Let us first estimate the second term $\Delta H_{\partial\Omega}^{(1)}$. 
For this purpose, we decompose $\tilde{\Omega}$ into three parts as 
$$
\tilde{\Omega}=\tilde{\Omega}^{(-)}\cup\tilde{\Omega}^{(0)}\cup\tilde{\Omega}^{(+)}
$$
with 
$$
\tilde{\Omega}^{(-)}:=\tilde{D}_{-\tilde{L}}\cup\cdots\cup\tilde{D}_{-1},
$$
$$
\tilde{\Omega}^{(0)}:=\tilde{D}_0\cup\tilde{D}_1\cup\cdots\cup\tilde{D}_L,
$$
and
$$
\tilde{\Omega}^{(+)}:=\tilde{D}_{L+1}\cup\cdots\cup\tilde{D}_{\tilde{L}}.
$$
Using this decomposition, the operator $\Delta H_{\partial\Omega}^{(1)}$ further decompose into three 
parts as 
$$
\Delta H_{\partial\Omega}^{(1)}=\Delta H_{\partial\Omega}^{(1,-)} +
\Delta H_{\partial\Omega}^{(1,0)}+\Delta H_{\partial\Omega}^{(1,+)},
$$
where 
$$
\Delta H_{\partial\Omega}^{(1,-)}:=\sum_{\alpha,\beta}\sum_{x\in\Omega}
\sum_{y\in\tilde{\Omega}^{(-)}}t_{x,y}^{\alpha,\beta}
\left\{\exp[-i\theta_\pm(x)]-1\right\}c_{x,\alpha}^\dagger c_{y,\beta},
$$
$$
\Delta H_{\partial\Omega}^{(1,0)}:=
\sum_{\alpha,\beta}\sum_{x\in\Omega}
\sum_{y\in\tilde{\Omega}^{(0)}\backslash\Omega}t_{x,y}^{\alpha,\beta}
\left\{\exp[-i\theta_\pm(x)]-1\right\}c_{x,\alpha}^\dagger c_{y,\beta},
$$
and 
$$
\Delta H_{\partial\Omega}^{(1,+)}:=\sum_{\alpha,\beta}\sum_{x\in\Omega}
\sum_{y\in\tilde{\Omega}^{(+)}}t_{x,y}^{\alpha,\beta}
\left\{\exp[-i\theta_\pm(x)]-1\right\}c_{x,\alpha}^\dagger c_{y,\beta}.
$$
We assume that all of the discs $D_j$ satisfy $|D_j|={\cal O}(R^{d-1})$, where $R$ is the radius of the discs. 
Then, the volume $|\Omega|$ of the region $\Omega$ satisfies $|\Omega|={\cal O}(R^{d-1}\times L)$. 
We also assume that the area of the side of the cylindrical region $\Omega$ 
satisfies $|\partial\Omega\backslash(D_0\cup D_L)|={\cal O}(R^{d-2}\times L)$.  
Since the hopping range is of finite, we have 
$$
\left\Vert\Delta H_{\partial\Omega}^{(1,\pm)}\right\Vert\le {\rm Const.}\times \frac{R^{d-1}}{L} 
$$
{from} the definition (\ref{theta}) of $\theta_\pm(x)$. In the same way, one has 
$$
\left\Vert\Delta H_{\partial\Omega}^{(1,0)}\right\Vert 
\le {\rm Const.}\times R^{d-2} L. 
$$
Combining these, we obtain 
\begin{equation}
\label{estDeltaH1}
\frac{L}{|\Omega|}\left\Vert\Delta H_{\partial\Omega}^{(1)}\right\Vert
\le {\rm Const.}\times\frac{1}{L}+{\rm Const.}\times\frac{L}{R}.
\end{equation}
In the same way, we also have 
\begin{equation}
\label{estDeltaH2}
\frac{L}{|\Omega|}\left\Vert\Delta H_{\partial\Omega}^{(2)}\right\Vert
\le {\rm Const.}\times\frac{1}{L}+{\rm Const.}\times\frac{L}{R}.
\end{equation}

Next, consider $\Delta H_\Omega$. From (\ref{theta}) and (\ref{def:ell}), one has 
$$
\Delta H_\Omega=\pm\frac{2\pi}{L}\sum_{\alpha,\beta}\sum_{x,y\in\Omega:\ell(x)<\ell(y)}
i[\ell(y)-\ell(x)]\left(t_{x,y}^{\alpha,\beta}c_{x,\alpha}^\dagger c_{y,\beta}-t_{y,x}^{\beta,\alpha}
c_{y,\beta}^\dagger c_{x,\alpha}\right)+{\cal O}(|\Omega|\times L^{-2}).
$$
Write $k=\ell(y)-\ell(x)$ for the factor in the summand in the right-hand side. 
In the expression (\ref{JOmega}) of the current $J_\Omega$ on 
the region $\Omega$, the nonvanishing contributions for $x,y$ satisfying $k=\ell(y)-\ell(x)$ 
appear $k$ times in the sums. From these observations, we have  
$$
\frac{1}{2\pi}\frac{L}{|\Omega|}\Delta H_\Omega \mp J_\Omega 
={\cal O}(L^{-1})+{\cal O}(L/R).
$$
Combining this, (\ref{DeltaE}), (\ref{diffH}), (\ref{estDeltaH1}) and (\ref{estDeltaH2}), 
we obtain 
$$
\pm \lim_{L\nearrow\infty}\lim_{R\nearrow\infty}\omega_0(J_\Omega)\ge 0, 
$$
where $\omega_0$ is the infinite-volume ground state, and the double limit is taken 
so that $L/R\rightarrow 0$ in the limit. 
This implies the vanishing of the bulk persistent current, i.e., (\ref{zeroAvJ}). 
\end{proof}

%%%%%%%%%%%%%%%%%%%%%%%%%%%%%%%%%%%%%%%%%%%%%%%%%%
\Section{Absence of $U(1)$ symmetry breaking}
\label{sec:Elitzur}

In this section, we extend Elitzur's theorem to fermionic models.
Consider the Hamiltonian $H_\Lambda$ of (\ref{HamLambda}) on a $d$-dimensional finite lattice $\Lambda\subset \ze^d$. 
For simplicity, we assume that the Hamiltonian $H_\Lambda$ contains only the nearest neighbor hopping, 
and that the fermions have only spin-1/2 as the internal degree of freedom, i.e., we consider usual electrons. 

%%%%%%%%%%%%%%%%%%%%%%%%%%%%%%%%%%%
\subsection{Annealed gauge fields}

In order to take into account the fluctuation of an electromagnetic field, 
we introduce a $U(1)$ gauge field $A$ as follows: For each nearest neighbor pair $\langle x,y\rangle$ of 
sites $x,y\in\Lambda$, 
the gauge field $A_{x,y}$ takes the value $A_{x,y}\in\re\ \mbox{mod}\ 2\pi$, and satisfies the conditions, 
$$
A_{y,x}=-A_{x,y}\ \mbox{mod}\ 2\pi.
$$
The hopping amplitudes of the Hamiltonian $H_\Lambda$ of (\ref{HamLambda}) is replaced with 
$$
{\tilde t}_{x,y}^{\alpha,\beta}(A):=t_{x,y}^{\alpha,\beta}e^{iA_{x,y}}
$$
for each nearest neighbor pair $\langle x,y\rangle$ of sites. Then, the Hamiltonian of electrons coupled 
to the gauge field $A$ is given by  
\begin{eqnarray}
{\tilde H}_\Lambda(A)
&:=&\sum_{x,y\in\Lambda}\sum_{\alpha,\beta}{\tilde t}_{x,y}^{\alpha,\beta}(A)c_{x,\alpha}^\dagger c_{y,\beta}\nonumber\\
&+&\sum_{I\ge 1}\sum_{x_1,\sigma_1}\sum_{x_2,\sigma_2}\cdots\sum_{x_I,\sigma_I}
W_{x_1,\sigma_1;x_2,\sigma_2;\ldots;x_I,\sigma_I}n_{x_1,\sigma_1}n_{x_2,\sigma_2}\cdots n_{x_I,\sigma_I}.
\label{HamA}
\end{eqnarray}
As a local order parameter to detect the $U(1)$ symmetry breaking, 
we consider the Cooper pair $c_{u,\uparrow}c_{v,\downarrow}$ of electrons for fixed two sites $u,v$, 
where we assume that the site $v$ is in a neighborhood $\mathcal{N}_u$ of the site $u$. 
The total energy containing the energy of the gauge fields and the terms of the symmetry breaking fields is  
\begin{equation}
\label{energy}
\mathcal{H}_\Lambda^{(N)}(A):={\tilde H}_\Lambda(A)-\mu N_\Lambda-\kappa\sum_p\cos B_p
-h H_{{\rm ext},\Lambda}-\tilde{h}\tilde{H}_{{\rm ext},\Lambda}(A),
\end{equation}
where $N_\Lambda$ is the total number operator of electrons with the chemical potential $\mu$, i.e., 
$$
N_\Lambda:=\sum_{\sigma}\sum_{x\in\Lambda}n_{x,\sigma},
$$
and $B_p$ is the magnetic flux through the plaquette $p$ (unit square cell); 
the Hamiltonian $H_{{\rm ext},\Lambda}$ of the symmetry breaking field 
is given by the sum of the local order parameters as   
$$
H_{{\rm ext},\Lambda}:=\sum_{x\in\Lambda}\sum_{y\in\mathcal{N}_x}\left(c_{x,\uparrow}c_{y,\downarrow}
+c_{y,\downarrow}^\dagger c_{x,\uparrow}^\dagger \right),
$$
and the last term in the energy (\ref{energy}) is also the Hamiltonian of the symmetry breaking field,  
\begin{equation}
\tilde{H}_{{\rm ext},\Lambda}(A):=\sum_{b\in\mathcal{B}}\cos A_b,
\label{tildeHext} 
\end{equation}
for the gauge field $A$. Here, $\mathcal{B}$ is the set of the bonds 
(the nearest neighbor pairs of sites), and we have written $A_b=A_{x,x+e_i}$ with the unit vector $e_i$ 
in the $i$-th direction. 
The three parameters, $\kappa, h$ and $\tilde{h}$, are taken to be positive. 

The expectation value of the Cooper pair at an inverse temperature $\beta$ is given by 
\begin{equation}
\label{Cooperexp}
\langle c_{u,\uparrow}c_{v,\downarrow}\rangle_\Lambda(\beta,h,\tilde{h},\kappa):=
\frac{1}{Z_\Lambda(\beta)}\int_{-\pi}^{\pi}\prod_{b\in\mathcal{B}}dA_b\; 
{\rm Tr}\; c_{u,\uparrow}c_{v,\downarrow}\exp[-\beta \mathcal{H}_\Lambda^{(N)}(A)], 
\end{equation}
where $Z_\Lambda(\beta)$ is the partition function. In the following, we take 
the limit $\kappa\uparrow\infty$ so that all the magnetic flux $B_p$ through the plaquette $p$ are 
vanishing. We stress that our result still holds for a finite $\kappa$. But we impose 
this strong condition because a nonvanishing magnetic flux generally is believed to suppress superconducting states.
In the infinite limit $\kappa\uparrow\infty$, the gauge fixing degree of freedom still remains. 
Therefore, in order to lift the gauge fixing degree of the freedom, we have introduced 
the Hamiltonian $\tilde{H}_{{\rm ext},\Lambda}(A)$ of (\ref{tildeHext}) which 
prefers the gauge fixing $A_b=0$ for all the bonds $b$. 
  
Let us consider the spontaneous magnetization for the Cooper pair in the infinite-volume limit and 
the zero magnetic field limit. Namely, we define   
\begin{equation}
\label{Cooperinf}
\langle c_{u,\uparrow}c_{v,\downarrow}\rangle(\beta):=
\lim_{\tilde{h}\downarrow 0}\lim_{h\downarrow 0}
\lim_{\Lambda\nearrow \ze^d}\lim_{\kappa\nearrow \infty}
  \langle c_{u,\uparrow}c_{v,\downarrow}\rangle_\Lambda(\beta,h,\tilde{h},\kappa). 
\end{equation}
More precisely, the left-hand side denotes all of the accumulation points in the limit. 

We have: 

\begin{theorem}
\label{thm:zeroCooper} 
The spontaneous magnetization for the Cooper pair is vanishing in the infinite-volume limit and 
in the zero magnetic field, i.e., 
\begin{equation}
\label{Cooperzero}
\langle c_{u,\uparrow}c_{v,\downarrow}\rangle(\beta)=0, 
\end{equation}
for any inverse temperatures $\beta$. 
\end{theorem}
\smallskip

\noindent
{\it Remark:} (i) Instead of the Cooper pair amplitude, 
the statement of Theorem~\ref{thm:zeroCooper} holds for any local observable 
which transforms nontrivially under the local gauge transformations. 
\smallskip

\noindent (ii) Consider the situation without the symmetry breaking fields. 
But, instead of the symmetry breaking fields, we introduce gauge fixing terms into the Hamiltonian as 
in Kennedy and King \cite{KK} and Borgs and Nill \cite{BN2,BN1}. 
For the Higgs models, they proved that the two-point correlations for the Higgs fields 
do not exhibit long-range order in dimensions $d\le 4$ except for the Landau gauge.  
In the Landau gauge, Kennedy and King \cite{KK,BN3} 
proved that the noncompact $U(1)$ Higgs model has a phase transition in dimensions $d\ge 3$. 
However, when there appears the remaining gauge degree of freedom which is called the Gribov ambiguity, 
the two-point correlations vanish at noncoinciding points even in the Landau gauge 
as Borgs and Nill \cite{BN2,BN1} pointed out. 

In the same situation for the present lattice fermion systems, 
we can prove that the two-point Cooper pair correlations do not show long-range order 
in spatial dimensions $d\le 4$ except for the Landau gauge. Further, 
if the corresponding Gribov ambiguity appears, then the two-point correlations vanish at noncoinciding points 
even in the Landau gauge. We will publish the details elsewhere.
\medskip

\begin{proof}{Proof}
We introduce the unitary transformation with twisting angle $\theta\in\re\ \mbox{mod}\ 2\pi$ 
at the site $u$ for the Cooper pair $c_{u,\uparrow}c_{v,\downarrow}$ as
$$
U_{u,\uparrow}(\theta):=\exp[i\theta n_{u,\uparrow}].
$$
Since one has 
$$
(U_{u,\uparrow}(\theta))^\ast c_{u,\uparrow}U_{u,\uparrow}(\theta)=e^{i\theta}c_{u,\uparrow}, 
$$
the hopping amplitudes in the Hamiltonian $\tilde{H}_\Lambda(A)$ of (\ref{HamA}) 
for electrons change under the transformation as 
$$
\tilde{t}_{x,y}^{\alpha,\beta}(A)\rightarrow \tilde{t}_{x,y}^{\alpha,\beta}(A'),
$$
where 
$$
A'_{x,y}=A_{x,y}-\theta \quad \mbox{for} \ x=u \ \mbox{and}\ y=u+e_i,
$$
$$
A'_{y,x}=A_{y,x}+\theta \quad \mbox{for} \ x=u \ \mbox{and}\ y=u-e_i,
$$
and $A'_{x,y}=A_{x,y}$, otherwise. Therefore, the energy operator of (\ref{energy}) is transformed as 
\begin{eqnarray}
((U_{u,\uparrow}(\theta))^\ast\mathcal{H}_\Lambda^{(N)}U_{u,\uparrow}(\theta))(A)
&=&{\tilde H}_\Lambda(A')-\mu N_\Lambda-\kappa\sum_p\cos B_p \nonumber \\
&-&h (U_{u,\uparrow}(\theta))^\ast H_{{\rm ext},\Lambda}U_{u,\uparrow}(\theta)-\tilde{h}\tilde{H}_{{\rm ext},\Lambda}(A).
\nonumber
\end{eqnarray}
Write  
$$
\Delta H_{\rm ext}:=(U_{u,\uparrow}(\theta))^\ast H_{{\rm ext},\Lambda}U_{u,\uparrow}(\theta)
-H_{{\rm ext},\Lambda}
$$
and 
$$
\Delta \tilde{H}_{\rm ext}:=\tilde{H}_{{\rm ext},\Lambda}(A)
-\tilde{H}_{{\rm ext},\Lambda}(A').
$$
Clearly, both of these are local. Then, we have 
$$
((U_{u,\uparrow}(\theta))^\ast\mathcal{H}_\Lambda^{(N)}U_{u,\uparrow}(\theta))(A)
=\mathcal{H}_\Lambda^{(N)}(A')-h\Delta H_{\rm ext}-\tilde{h}\Delta \tilde{H}_{\rm ext}. 
$$
Using this unitary transformation and changing the variables in the integral for 
the expectation value of (\ref{Cooperexp}), we have 
\begin{equation}
\label{Cooperxexp'}
\langle c_{u,\uparrow}c_{v,\downarrow}\rangle_\Lambda(\beta,h,\tilde{h},\kappa)=
\frac{e^{i\theta}}{Z_\Lambda(\beta)}\int_{-\pi}^{\pi}\prod_{b\in\mathcal{B}}dA_b'\; 
{\rm Tr}\; c_{u,\uparrow}c_{v,\downarrow}\exp[-\beta \mathcal{H}_\Lambda^{(N)}(A')
-\beta \Delta\mathcal{H}_{\rm ext}], 
\end{equation}
where we have written 
$$
\Delta\mathcal{H}_{\rm ext}:=h\Delta H_{{\rm ext}}+\tilde{h}\Delta \tilde{H}_{{\rm ext}}
$$
for short. 

Let $O$ be an observable. Note that 
\begin{eqnarray}
\Delta N(O)&:=&{\rm Tr}\; O\exp[-\beta \mathcal{H}_\Lambda^{(N)}(A')
-\beta \Delta\mathcal{H}_{\rm ext}]-{\rm Tr}\; O\exp[-\beta \mathcal{H}_\Lambda^{(N)}(A')]\nonumber \\
&=&\int_0^1 d\lambda \frac{d}{d\lambda}{\rm Tr}\; O\exp[-\beta \mathcal{H}_\Lambda^{(N)}(A')
-\lambda\beta \Delta\mathcal{H}_{\rm ext}]\nonumber \\
&=&(-\beta)\int_0^1 d\lambda(O,\Delta\mathcal{H}_{\rm ext})
{\rm Tr}\; \exp[-\beta \mathcal{H}_\Lambda^{(N)}(A')
-\lambda\beta \Delta\mathcal{H}_{\rm ext}],
\label{defDeltaNO}
\end{eqnarray}
where $(O_1,O_2)$ is the Duhamel two-point function \cite{Roepstrff,DLS} which is given  by 
$$
(O_1,O_2):=\frac{1}{Z}\int_0^1 dz\; {\rm Tr}\; O_1e^{-z\beta\mathcal{H}}O_2e^{-(1-z)\beta\mathcal{H}}
$$
for the observables $O_1, O_2$. 
In order to evaluate the right-hand side of (\ref{defDeltaNO}), let us consider  
\begin{eqnarray}
& &\log{\rm Tr}\; \exp[-\beta \mathcal{H}_\Lambda^{(N)}(A')-\lambda\beta \Delta\mathcal{H}_{\rm ext}]
-\log {\rm Tr}\; \exp[-\beta \mathcal{H}_\Lambda^{(N)}(A')]\nonumber \\ 
&=&\int_0^1 d\lambda' \frac{d}{d\lambda'}\log{\rm Tr}\; \exp[-\beta \mathcal{H}_\Lambda^{(N)}(A')
-\lambda'\lambda\beta \Delta\mathcal{H}_{\rm ext}] \nonumber \\
&=&(-\lambda\beta)\int_0^1 d\lambda' \frac{{\rm Tr}\; \Delta \mathcal{H}_{\rm ext}\exp[-\beta \mathcal{H}_\Lambda^{(N)}(A')
-\lambda'\lambda\beta \Delta\mathcal{H}_{\rm ext}]}{{\rm Tr}\; \exp[-\beta \mathcal{H}_\Lambda^{(N)}(A')
-\lambda'\lambda\beta \Delta\mathcal{H}_{\rm ext}]}.
\end{eqnarray}
Therefore, one has 
$$
\left|\log \frac{{\rm Tr}\; \exp[-\beta \mathcal{H}_\Lambda^{(N)}(A')
-\lambda\beta \Delta\mathcal{H}_{\rm ext}]}{{\rm Tr}\; \exp[-\beta \mathcal{H}_\Lambda^{(N)}(A')]}\right|\le |\lambda|\beta 
\left\Vert\Delta \mathcal{H}_{\rm ext}\right\Vert. 
$$
This implies  
\begin{equation}
\label{estZ}
e^{-|\lambda|\beta\Vert\Delta \mathcal{H}_{\rm ext}\Vert}
\le \frac{{\rm Tr}\; \exp[-\beta \mathcal{H}_\Lambda^{(N)}(A')
-\lambda\beta \Delta\mathcal{H}_{\rm ext}]}{{\rm Tr}\; \exp[-\beta \mathcal{H}_\Lambda^{(N)}(A')]}
\le e^{|\lambda|\beta\Vert\Delta \mathcal{H}_{\rm ext}\Vert}. 
\end{equation}
Note that the Duhamel two-point function satisfies \cite{Roepstrff,DLS}
$$
|(O_1,O_2)|\le\Vert O_1\Vert\; \Vert O_2\Vert. 
$$
Using this and the inequality (\ref{estZ}), $\Delta N(O)$ of (\ref{defDeltaNO}) can be estimated as 
\begin{eqnarray}
|\Delta N(O)|&\le& \beta \Vert O\Vert\; \Vert \Delta\mathcal{H}_{\rm ext}\Vert 
e^{\beta(ah+\tilde{a}\tilde{h})}\;{\rm Tr}\; \exp[-\beta \mathcal{H}_\Lambda^{(N)}(A')]\nonumber \\
&\le &\beta (ah+\tilde{a}\tilde{h})\Vert O\Vert\;  
e^{\beta(ah+\tilde{a}\tilde{h})}\;{\rm Tr}\; \exp[-\beta \mathcal{H}_\Lambda^{(N)}(A')],
\label{estDeltaNO}
\end{eqnarray}
where $a$ and $\tilde{a}$ are some positive constant. 

By using $\Delta N(c_{u,\uparrow}c_{v,\downarrow})$ of (\ref{defDeltaNO}) 
for the right-hand side of (\ref{Cooperxexp'}), we obtain 
\begin{equation}
\langle c_{u,\uparrow}c_{v,\downarrow}\rangle_\Lambda(\beta,h,\tilde{h},\kappa)
=e^{i\theta}
\langle c_{u,\uparrow}c_{v,\downarrow}\rangle_\Lambda(\beta,h,\tilde{h},\kappa)
+\frac{e^{i\theta}}{Z_\Lambda(\beta)}\int_{-\pi}^{\pi}\prod_{b\in\mathcal{B}}dA_b' \Delta N(c_{u,\uparrow}c_{v,\downarrow}).
\label{Cooperest3}
\end{equation}
{From} the inequality (\ref{estDeltaNO}), the second term in the right-hand side is estimated as  
\begin{eqnarray}
& &\left|\frac{1}{Z_\Lambda(\beta)}\int_{-\pi}^{\pi}\prod_{b\in\mathcal{B}}dA_b' 
\Delta N(c_{u,\uparrow}c_{v,\downarrow})\right|\nonumber \\
&\le& \frac{1}{Z_\Lambda(\beta)}\int_{-\pi}^{\pi}\prod_{b\in\mathcal{B}}dA_b'\; 
\beta (ah+\tilde{a}\tilde{h})e^{\beta(ah+\tilde{a}\tilde{h})}\Vert c_{u,\uparrow}c_{v,\downarrow}\Vert\;
{\rm Tr}\; \exp[-\beta\mathcal{H}_\Lambda^{(N)}(A')]\nonumber \\
&\le & \beta (ah+\tilde{a}\tilde{h})e^{\beta(ah+\tilde{a}\tilde{h})}\Vert c_{u,\uparrow}c_{v,\downarrow}\Vert.
\label{errorCooper}
\end{eqnarray}
Then, from (\ref{Cooperinf}), (\ref{Cooperest3}) and (\ref{errorCooper}), we obtain 
$$
\langle c_{u,\uparrow}c_{v,\downarrow}\rangle(\beta)
=e^{i\theta}\langle c_{u,\uparrow}c_{v,\downarrow}\rangle(\beta).
$$
This implies the desired result (\ref{Cooperzero}).
\end{proof}

%%%%%%%%%%%%%%%%%%%%%%%%%%%%%%%%%%%
\subsection{Quenched gauge fields}

Next, we extend Elitzur's theorem to the present models whose configurations of gauge fields are quenched. 
For a given gauge fields $A$, let us consider the Hamiltonian, 
\begin{equation}
\hat{\mathcal{H}}_\Lambda^{(N)}(A)={\tilde H}_\Lambda(A)-\mu N_\Lambda
-h H_{{\rm ext},\Lambda}
\end{equation}
with the Hamiltonian of the symmetry breaking field, 
$$
H_{{\rm ext},\Lambda}=\sum_{x\in\Lambda}\sum_{y\in\mathcal{N}_x}\left(c_{x,\uparrow}c_{y,\downarrow}
+c_{y,\downarrow}^\dagger c_{x,\uparrow}^\dagger \right).
$$
Then, the expectation value is given by 
\begin{equation}
\langle \cdots\rangle_\Lambda(\beta,h,A)=
\frac{1}{Z_\Lambda(\beta,A)}
{\rm Tr}(\cdots)\exp[-\beta \hat{\mathcal{H}}_\Lambda^{(N)}(A)], 
\end{equation}
where $Z_\Lambda(\beta,A)$ is the partition function. As an example, 
we consider the expectation value of the Cooper pair 
$\langle c_{u,\uparrow}c_{v,\downarrow}\rangle_\Lambda(\beta,h,A)$. Here, we have not introduced 
the Hamiltonian of the symmetry breaking field (\ref{tildeHext}) because the term plays no role in 
the expectation value. The average with respect to the gauge fields $A$ is given by 
\begin{equation}
\av_{\Lambda,\kappa}\left[\cdots\right]:=\int_{-\pi}^{\pi}\prod_{b\in\mathcal{B}}dA_b\;\rho_{\Lambda,\kappa}(A) 
\left(\cdots\right)
\end{equation}
with the probability density, 
$$
\rho_{\Lambda,\kappa}(A):=\frac{\exp[\kappa\sum_p\cos B_p]}{\int_{-\pi}^{\pi}\prod_{b\in\mathcal{B}}dA_b\;
\exp[\kappa\sum_p\cos B_p]}.
$$
In the limit $\kappa\nearrow\infty$, the zero magnetic field is realized. In the following, we take 
the limit as in the preceding section. 

We write 
\begin{equation}
\label{MLambda}
\mathcal{M}_\Lambda:=\frac{1}{|\Lambda|}\sum_{x\in\Lambda}\sum_{y\in\mathcal{N}_x}
\left(c_{x,\uparrow}c_{y,\downarrow}+c_{y,\downarrow}^\dagger c_{x,\uparrow}^\dagger\right),
\end{equation}
and 
$$
M(\beta,h,A):=\lim_{\Lambda\nearrow\ze^d} 
\langle\mathcal{M}_\Lambda\rangle_\Lambda(\beta,h,A).
$$
We define the spontaneous magnetization $M(\beta,A)$ and the long-range order $\sigma(\beta,A)$ as  
$$
M(\beta,A):=\lim_{h\downarrow 0}M(\beta,h,A)
$$
and
$$
\sigma(\beta,A):=\lim_{\Lambda\nearrow\ze^d} 
\sqrt{\langle(\mathcal{M}_\Lambda)^2\rangle_\Lambda(\beta,0,A)}. 
$$
Here, we consider all of the accumulation points again. 

We obtain:

\begin{theorem} 
\label{thm:zeroMsigma}
Both of 
the spontaneous magnetization $M(\beta,A)$ and the long-range order $\sigma(\beta,A)$ 
for the Cooper pair are vanishing for almost all gauge fields $A$.
\end{theorem}
\smallskip

\noindent
{\it Remark:} (i) Theorem~\ref{thm:zeroMsigma} does not exclude the possibility that 
the $U(1)$ symmetry breaking occurs for some particular gauge fixing. 
As mentioned in Remark~(ii) of Theorem~\ref{thm:zeroCooper}, Kennedy and King \cite{KK,BN3} 
showed that there appears $U(1)$ symmetry breaking in a noncompact $U(1)$ Higgs model in Landau gauge. 
However, there are some differences between noncompact and compact gauge theories. 
In fact, except for this special case which was treated by Kennedy and King, several cases in a certain gauge fixing 
show absence of the $U(1)$ symmetry breaking in Higgs models \cite{BN2,FMS,BN1}. 
\smallskip

\noindent
(ii) Remark~(i) of Theorem~\ref{thm:zeroCooper} holds for the present case, too. 
\medskip

\begin{proof}{Proof}
Using the gauge transformation in the same way as in the preceding subsection, we have 
\begin{equation}
\av_{\Lambda,\kappa}\left[\langle c_{u,\uparrow}c_{v,\downarrow}\rangle_\Lambda(\beta,h,\cdots)\right]
=e^{i\theta}\int_{-\pi}^{\pi}\prod_{b\in\mathcal{B}}dA_b'\>
\frac{\rho_{\Lambda,\kappa}(A')}{Z_\Lambda'(\beta,A')}
{\rm Tr}\> c_{u,\uparrow}c_{v,\downarrow}
\exp[-\beta (\hat{\mathcal{H}}_\Lambda^{(N)}(A')+\Delta{\mathcal{H}}_{\rm ext})],
\end{equation}
where the deformed partition function is given by 
$$
Z_\Lambda'(\beta,A')=
{\rm Tr}\> \exp[-\beta (\hat{\mathcal{H}}_\Lambda^{(N)}(A')+\Delta{\mathcal{H}}_{\rm ext})].
$$
and 
$$
\Delta\mathcal{H}_{\rm ext}=h\left(U_{u,\uparrow}(\theta)\right)^\ast H_{{\rm ext},\Lambda}
U_{u,\uparrow}(\theta)-hH_{{\rm ext},\Lambda}.
$$
This expectation value can be written as 
\begin{eqnarray}
& &\av_{\Lambda,\kappa}\left[\langle c_{u,\uparrow}c_{v,\downarrow}\rangle_\Lambda(\beta,h,\cdots)\right]\nonumber\\
&=&e^{i\theta}\int_{-\pi}^{\pi}\prod_{b\in\mathcal{B}}dA_b'\>
\frac{\rho_{\Lambda,\kappa}(A')}{Z_\Lambda'(\beta,A')}
\left\{{\rm Tr}\> c_{u,\uparrow}c_{v,\downarrow}
\exp[-\beta (\hat{\mathcal{H}}_\Lambda^{(N)}(A')+\Delta{\mathcal{H}}_{\rm ext})]\right.\nonumber\\
& &\qquad \qquad \qquad \quad \qquad \qquad\left.-{\rm Tr}\> c_{u,\uparrow}c_{v,\downarrow}
\exp[-\beta \hat{\mathcal{H}}_\Lambda^{(N)}(A')]\right\}\nonumber\\
&+&e^{i\theta}\int_{-\pi}^{\pi}\prod_{b\in\mathcal{B}}dA_b'\>
\frac{\rho_{\Lambda,\kappa}(A')}{Z_\Lambda(\beta,A')}\left[\frac{Z_\Lambda(\beta,A')}{Z_\Lambda'(\beta,A')}-1
\right]{\rm Tr}\> c_{u,\uparrow}c_{v,\downarrow}
\exp[-\beta \hat{\mathcal{H}}_\Lambda^{(N)}(A')]\nonumber\\
&+&e^{i\theta}
\av_{\Lambda,\kappa}\left[\langle c_{u,\uparrow}c_{v,\downarrow}\rangle_\Lambda(\beta,h,\cdots)\right].
\end{eqnarray}
The first and the second terms in the right-hand side can be evaluated in the same way as in (\ref{estDeltaNO}) 
and (\ref{estZ}), respectively. In consequence, we obtain 
\begin{equation}
\label{avexpCooper}
\left|\av_{\Lambda,\kappa}\left[\langle c_{u,\uparrow}c_{v,\downarrow}\rangle_\Lambda(\beta,h,\cdots)\right]\right|
\le {\rm Const.}\frac{\beta h}{|1-e^{i\theta}|}e^{\beta a h}
\end{equation} 
with some positive constant $a$. 
This implies 
$$
\lim_{h\downarrow 0}\lim_{\Lambda\nearrow \ze^d}\lim_{\kappa\uparrow\infty}
\av_{\Lambda,\kappa}\left[\langle c_{u,\uparrow}c_{v,\downarrow}\rangle_\Lambda(\beta,h,\cdots)\right]=0
$$
for any pair of two sites $u,v$. Similarly, for $\mathcal{M}_\Lambda$ of (\ref{MLambda}), we have 
\begin{eqnarray}
& &\hspace{-1.7cm}
\left|\av_{\Lambda,\kappa}\left[\langle \mathcal{M}_\Lambda\rangle_\Lambda(\beta,h,\cdots)\right]\right|\nonumber\\
&\le& \frac{1}{|\Lambda|}\sum_{x\in\Lambda}\sum_{y\in\mathcal{N}_x}\left\{
\Bigl|\av_{\Lambda,\kappa}\left[\langle c_{x,\uparrow}c_{y,\downarrow}\rangle_\Lambda(\beta,h,\cdots)\right]\Bigr|
+\Bigl|\av_{\Lambda,\kappa}\bigl[\langle 
c_{y,\downarrow}^\dagger c_{x,\uparrow}^\dagger \rangle_\Lambda(\beta,h,\cdots)\bigr]\Bigr|\right\} \nonumber \\
&\le& {\rm Const.}\frac{\beta h}{|1-e^{i\theta}|}e^{\beta a h} \nonumber
\end{eqnarray}
by using inequalities similar to the above inequality (\ref{avexpCooper}). Immediately, 
\begin{equation}
\lim_{h\downarrow 0}\lim_{\Lambda\nearrow \ze^d}\lim_{\kappa\uparrow\infty}
\av_{\Lambda,\kappa}\left[\langle \mathcal{M}_\Lambda\rangle_\Lambda(\beta,h,\cdots)\right]=0.
\label{EcalM=0}
\end{equation}

Note that the global $U(1)$ transformation contains the reversal of the orientation of 
the external symmetry breaking field. Under this transformation, the free energy is invariant. 
Combining this fact with the concavity of the free energy, one can show the positivity of the magnetization, i.e., 
$$
\langle\mathcal{M}_\Lambda\rangle_\Lambda(\beta,h,A)\ge 0
$$ 
for $h\ge 0$. Let $\varepsilon$ be a small positive number, and 
consider the probability that the events satisfying 
$\langle\mathcal{M}_\Lambda\rangle_\Lambda(\beta,h,A)\ge \varepsilon$ occur. Then, Markov's inequality yields  
$$
\av_{\Lambda,\kappa}\left[\langle\mathcal{M}_\Lambda\rangle_\Lambda(\beta,h,\cdots)\right]
\ge \varepsilon {\rm Prob}\left[\langle\mathcal{M}_\Lambda\rangle_\Lambda(\beta,h,\cdots)\ge \varepsilon\right].
$$
{From} the above result (\ref{EcalM=0}), we obtain 
\begin{eqnarray}
0=\lim_{h\downarrow 0}\lim_{\Lambda\nearrow \ze^d}\lim_{\kappa\uparrow\infty}
\av_{\Lambda,\kappa}\left[\langle \mathcal{M}_\Lambda\rangle_\Lambda(\beta,h,\cdots)\right]
&\ge& \varepsilon \lim_{h\downarrow 0}\lim_{\Lambda\nearrow \ze^d}\lim_{\kappa\uparrow\infty}
{\rm Prob}\left[\langle\mathcal{M}_\Lambda\rangle_\Lambda(\beta,h,\cdots)\ge \varepsilon\right]\nonumber \\
&=&\varepsilon \lim_{h\downarrow 0}
{\rm Prob}\left[M(\beta,h,\cdots)\ge \varepsilon\right],
\end{eqnarray}
where ${\rm Prob}\left[M(\beta,h,\cdots)\ge \varepsilon\right]$ 
is the probability that the magnetization $M(\beta,h,A)$ in the infinite-volume limit satisfies 
$M(\beta,h,A)\ge\varepsilon$. This implies that the probability that the spontaneous magnetization $M(\beta,A)$ 
is greater than or equal to $\varepsilon$ is vanishing. Since the small positive number $\varepsilon$ is arbitrary, 
we obtain that the spontaneous magnetization $M(\beta,A)$ is vanishing with probability one.   

In order to show that the long-range order $\sigma(\beta,A)$ for the Cooper pair is vanishing, we recall 
the relation \cite{KomaTasaki} between the spontaneous magnetization and the long-range order as 
\begin{equation}
\label{KomaTasakiresult}
M(\beta,A)\ge \sigma(\beta,A)\ge 0.
\end{equation}
In their derivation, they relied on the existence of the thermodynamic limit of the free energy per volume. 
However, for an inhomogeneous system such as the present system with a fixed configuration of random gauge fields, 
the thermodynamic limit of the free energy per volume may not be unique. 
Therefore, we consider all of the accumulation points 
in the limit, by relying on the boundedness of the free energy per volume.   
Combining the inequality (\ref{KomaTasakiresult}) with the vanishing of the spontaneous magnetization $M(\beta,A)$, 
we obtain the desired result, i.e., the vanishing of the long-range order with probability one.  
\end{proof}

%%%%%%%%%%%%%%%%%%%%%%%%%%%%%%%%%%%%%%%%%%%%%%%%%%%%%%%%%%%%%%%%%%%%%%%%%%%
\bigskip

\noindent
{\bf Acknowledgements:} We would like to thank Peter Fulde, Hosho Katsura, Masaaki Shimozawa, 
Hal Tasaki and Masafumi Udagawa for helpful discussions. 
YT was partly supported by JSPS/MEXT Grant-in-Aid for Scientific
Research (Grant No. 26800177) and by Grant-in-Aid for
Program for Advancing Strategic International Networks to
Accelerate the Circulation of Talented Researchers (Grant No. R2604) ``TopoNet." 
%%%%%%%%%%%%%%%%%%%%%%%%%%%%%%%%%%%%%%%%%%%%%%%%%%%%%%%%%%%%%
%\newpage

\end{document}